\documentclass[twocolumn,aps,prb,10pt,floatfix]{revtex4-1}
\usepackage{amsmath} % For pmatrix
\usepackage{bm} % Bold Greek symbols
\usepackage{graphicx}
\usepackage[utf8]{inputenc}
\usepackage{wasysym} % For the symbol greater similar to
\usepackage[unicode=true]{hyperref}
\usepackage[caption=false]{subfig}
\usepackage{xcolor}

\begin{document}

\title{Electronic Ground State in Bilayer Graphene with Realistic Coulomb Interactions}

\author{Jia Ning Leaw}
\affiliation{Department of Physics and Centre for Advanced 2D Materials, National
University of Singapore, 6 Science Drive 2, 117546, Singapore}

\author{Ho-Kin Tang}
\affiliation{Department of Physics and Centre for Advanced 2D Materials, National
University of Singapore, 6 Science Drive 2, 117546, Singapore}

\author{Pinaki Sengupta}
\affiliation{{Centre for Advanced 2D Materials, National Univeristy of Singapore, 6 Science Drive 2, 117546 Singapore.}}
\affiliation{{School of Physical and Mathematical Sciences, Nanyang Technological University, 21 Nanyang Link, 637371 Singapore.}}

\author{Fakher F. Assaad}
\affiliation{{Institut f\"ur Theoretische Physik und Astrophysik, Universit\"at W\"urzburg, Am Hubland, D-97074 W\"urzburg, Germany.}}

\author{Igor F. Herbut}
\affiliation{{Department of Physics, Simon Fraser University, Burnaby, British Columbia V5A 1S6, Canada.}}

\author{Shaffique Adam}
%\email{shaffique.adam@yale-nus.edu.sg}
\affiliation{Department of Physics and Centre for Advanced 2D Materials, National
University of Singapore, 6 Science Drive 2, 117546, Singapore}
\affiliation{Yale-NUS College, 16 College Avenue West, 138614, Singapore}

\begin{abstract}
Both insulating and conducting electronic behaviors have been experimentally seen in clean bilayer graphene samples at low temperature, and there is still no consensus on the nature of the interacting ground state at half-filling and in the absence of a magnetic field.  Theoretically, several possibilities for the insulating ground states have been predicted for weak interaction strength.  However, a recent renormalization-group calculation on a Hubbard model for charge-neutral bilayer graphene with short-range interactions suggests the emergence of low-energy Dirac fermions that would stabilize the metallic phase for weak interactions.  Using a non-perturbative projective quantum Monte Carlo, we calculate the ground state for bilayer graphene using a realistic model for the Coulomb interaction that includes both short-range and long-range contributions.  We find that a finite critical onsite interaction is needed to gap bilayer graphene, thereby confirming the Hubbard model expectations even in the presence of a long-range Coulomb potential, in agreement with our theoretical renormalization group analysis. In addition, we also find that the critical onsite interactions necessary to destabilize the metallic ground state decreases with increasing interlayer coupling.

\end{abstract}
\maketitle 
\section{Introduction}

Stacking one layer of graphene on top of another dramatically changes
the dispersion. In the simplest consideration
where, in addition to the intra-plane hopping, only the hopping between
the carbon atoms that are placed directly on top of the other is allowed,
the non-interacting low-energy dispersion changes from linear to quadratic.  Although the stacking does not
directly gap out the bands, it does enhance the density of states
at the band-touching point, which increases quantum fluctuations and the 
likelihood of correlated ground states.  Moreover, onsite electron interactions are marginally relevant at half-filling, i.e. arbitrarily weak interactions will lead to an instability of the semi-metallic phase~\cite{Sun2009}. Bilayer graphene has additional degrees of freedom, which include spins, sublattices, layers and valleys, which has led to several theoretical predictions for different competing insulating phases including symmetry breaking of either real spin or pseudospin degrees
of freedom~\citep{Castro2008,Min2008,Zhang2010,Zhang2011,Jung2011,Zhang2012}, quantum anomalous Hall states~\citep{Nandkishore2010c},
nematic states~\citep{Vafek2010,Vafek2010a,Lemonik2010,Lemonik2012} and canted
antiferromagnetic states~\citep{Kharitonov2012}.  All of these theoretical works focused on the low-energy circular symmetric limit, where the dispersion is parabolic.

However, experiments remain inconclusive regarding what the ground state of bilayer
graphene at low temperature is, with zero electric and magnetic field,
and whether this system is conducting or gapped~\citep{Feldman2009,Weitz2010,Martin2010,Mayorov2011,VelascoJ2012,Freitag2012,Ulstrup2014}.
In particular, in the works by Bao et al. and Freitag et al.~\citep{Bao2012,Freitag2012},
both insulating and conducting phases were observed in similarly prepared samples.  The metallic samples were found to have low temperature conductivity of around $2$--$4\,e^{2}/h$, while that of the insulating samples was one order of magnitude lower.  This sample-dependent differences in the experimental data remains unresolved. 

In a recent theoretical development, Pujari et al. \citep{Pujari2016}
considered higher energies away from the band-touching point where
the Hamiltonian is no longer circular symmetric, and showed that a linear term can be generated by contact interactions.
Here, dimensional considerations show that the linear term once generated renders the onsite interaction irrelevant, and as such the system flows to a stable fixed point with Dirac cones.  Qualitatively, we can understand the stability of this Fermi liquid as resulting from the vanishing density of states of the emergent Dirac Fermions.  This conclusion was supported by quantum Monte Carlo simulations~\citep{Pujari2016} which show within this model, that for weak interactions, the system remains metallic, contradicting the previous expectation that interactions are marginally relevant.  Pujari et al. expected that their findings would not hold for realistic long-range Coulomb interactions since the same dimensional analysis shows that long-range interactions are relevant for this system, and as a result lead to an instability of the Fermi liquid for vanishing interactions.

However, as shown in Ref.~\onlinecite{Tang2018}, in the realistic system where the long-range
interactions are weaker than the contact interactions, the long-range
interactions will stabilize the Fermi liquid instead of destabilizing it. One way to understand this is that the long-range interactions
favor the charge-density-wave phase, which competes with the antiferromagnetic tendency 
favored by the contact interaction \cite{Hohenadler2014}. From the point of view of
renormalization group, the Coulomb interaction in 2D systems
has a non-analytical form of $2\pi e^{2}/\left|\mathbf{q}\right|$.
Since the Wilson's renormalization group cannot generate non-analytic terms, the Coulomb interaction vertex does not get directly renormalized~\citep{Gonzalez1994}; due to Ward
identity, however, the renormalization to the Coulomb interaction  is tied to the renormalization of the fermionic propagator. In the case of Dirac fermions where the fermionic propagator is described solely by the Fermi velocity $v_{F}$, the renormalization of the long-range interactions is captured by a single parameter $r_{s}=e^{2}/v_{F}$. Similarly, it  has also been shown for systems with parabolic dispersions that the renormalization group flow leads to a modification of the dynamical exponent,  which prevents the long-range interaction coupling constant to run away~\citep{Janssen2017}.
This can physically be understood as follows: Within the Hartree-Fock approximation,
the effective mass of the bilayer graphene is renormalized by long-range interactions to a reduced level~\citep{Borghi2009,ViolaKusminskiy2009}.
Since the strength of the long-range interactions is determined by the parameter $r_{s}=2me^{2}/\sqrt{\pi n}$ in bilayer graphene~\citep{DasSarma2011}, the reduced effective mass is equivalent to a weaker interaction.
Therefore, similarly to monolayer graphene where the enhanced Fermi
velocity reduces the long-range interaction strength, the downward renormalization of the  effective mass prevents the long-range interaction to escape to strong coupling.

The renormalization group arguments presented above are confirmed by large
scale unbiased qauntum Monte Carlo simulation of electrons on Bernal stacked bilayer
honeycomb lattice with on-site and long-range Coulomb interactions. Nevertheless, one
needs to be careful in interpreting the QMC results. 
%In addition to the renormalization group argument of Pujari
%et al., there is another reason why the phase transition occurs at
%a finite contact interaction in the quantum Monte Carlo simulations.
For non-interacting electrons on the bilayer honeycomb lattice with finite 
interlayer hopping $t_{\perp}$,
the electronic bands interpolate between a parabolic dispersion close
to the band touching point, and a linear dispersion at high momenta.
Such a crossover happens at the momentum $ka=t_{\perp}/\sqrt{3}t$~\citep{McCann2006}.
However, limited by finite system sizes, quantum Monte Carlo
simulations cannot probe momenta arbitrarily close to the band
touching point. For a system with $t_{\perp}=t$, the crossover
between a linear and parabolic dispersion occurs at $ka=1/\sqrt{3}$,
which is not much larger than the smallest momentum that can be accessed 
in simulations. 
The quantum Monte Carlo simulations with nominally
realistic values of inter-layer hopping are hence 
at the scales corresponding to the linear part of the band, where a large
finite critical on-site interaction is expected. Instead, one needs to use 
unrealistically large inter-layer hopping to probe the effects of on-site 
and long range interactions on the low-energy quadratic dispersion.

In this work, using numerically exact, projective quantum Monte Carlo \cite{Sugiyama86,Sorella1992,Hohenadler2014,Assaad08_rev}, we examine closely the
argument that a linear term is dynamically generated in bilayer graphene.
Our numerics suggests that the linear term is visibly generated only in the system
with $t_{\perp}\gg t$. In addition, our numerics support the conclusion that
the phase transition occurs at a finite onsite interaction strength
even in the presence of long-range interactions. The rest of the paper is organized as follows: In Sec.~\ref{subsec:Model}, we define the model
and the parameters, while in Sec.~\ref{subsec:Phase-Transition-of}
we present the quantum Monte Carlo results on the phase transition
of bilayer graphene for multiple values of the inter-layer hopping,
with $t \le t_\perp \le 10t$.  The renormalized spectrum is examined in Sec.~\ref{subsec:Spectrum-Renormalization}.
Finally, we discuss the relevance of our results to the realistic
system in Sec.~\ref{subsec:Discussion}.

\section{Model\label{subsec:Model}}

Our quantum Monte Carlo simulations assume  
single band interacting electrons on a bilayer honeycomb
lattice at half-filling.
%tight binding model of two honeycomb layers. 
In our model, the intralayer lattice vector
$\mathbf{a}_{1}=a\hat{x}$, $\mathbf{a}_{2}=a\hat{x}/2+\sqrt{3}a\hat{y}/2$
and the interlayer vector $\mathbf{a}_{3}=1.38a\hat{z}$, where $a$
is the lattice constant $2.46\,\text{\AA}$~\citep{Jung2014}. Each
unit cell consists of 4 sites $A_{1},B_{1},A_{2}$ and $B_{2}$, from
the two sub-lattices of the two layers, with $B_{1}$ sites sitting
on top of the $A_{2}$ sites. The Hamiltonian with both the onsite
Hubbard interactions and the long-range Coulomb interactions takes
the form
\begin{eqnarray*}
H & = & -t\sum_{\left\langle ij\right\rangle l\sigma}\hat{a}_{i\sigma l}^{\dagger}\hat{b}_{j\sigma l}-t_{\perp}\sum_{i}\hat{a}_{i\sigma2}^{\dagger}\hat{b}_{i\sigma1}+h.c.\\
 &  & \qquad+U\sum_{il}\left(\hat{a}_{i\uparrow l}^{\dagger}\hat{a}_{i\uparrow l}\hat{a}_{i\downarrow l}^{\dagger}\hat{a}_{i\downarrow l}+\hat{b}_{i\uparrow l}^{\dagger}\hat{b}_{i\uparrow l}\hat{b}_{i\downarrow l}^{\dagger}\hat{b}_{i\downarrow l}\right)\\ & & \qquad+\frac{1}{2}\sum_{i\neq j}V_{ij}\left(\hat{n}_{i}-1\right)\left(\hat{n}_{j}-1\right)
\end{eqnarray*}
The operator $a_{i\sigma l}^{\dagger}$ $\left(a_{i\sigma l}\right)$
acts on the atom in sub-lattice $A$ of the $l$-th layer, located
in the unit cell positioned at $\mathbf{r}_{i}$, to create (annihilate)
an electron of spin $\sigma=\uparrow\downarrow$. Similarly, $b_{i\sigma l}^{\dagger}$
and $b_{i\sigma l}$ act on sub-lattice $B$. The number operator
$\hat{n}_{i}=\sum_{\sigma}\hat{n}_{i\sigma}=\sum_{\sigma}\hat{c}_{i\sigma}^{\dagger}\hat{c}_{i\sigma}$
counts the number of electrons sitting at the atom $i$. For realistic
bilayer graphene, while the intra-layer hopping integral is $t=2.6\text{--}3.1\text{ eV}$,
the interlayer hopping $t_{\perp}$ is very small, usually about $10\%\text{--}15\%$
of $t$~\citep{Jung2014}. The interacting part of the Hamiltonian
consists of the onsite Hubbard interaction $U$ and long-range Coulomb
tail $V_{ij}=\gamma U/r_{ij}$, where $r_{ij}$ is the distance between
atom $i$ and atom $j$. The coupling constant for long-range interactions
$\alpha$ is related to the parameter in our model $\gamma$ as $\alpha=2U\gamma/3\sqrt{3}$.
We ignore other hopping integrals such as interlayer hopping between
atoms from the same sub-lattice which are negligible compared to $t$
and $t_{\perp}$. 

\section{Phase transition of bilayer graphene\label{subsec:Phase-Transition-of}}

\begin{figure}
\centering
\includegraphics[width=1\linewidth]{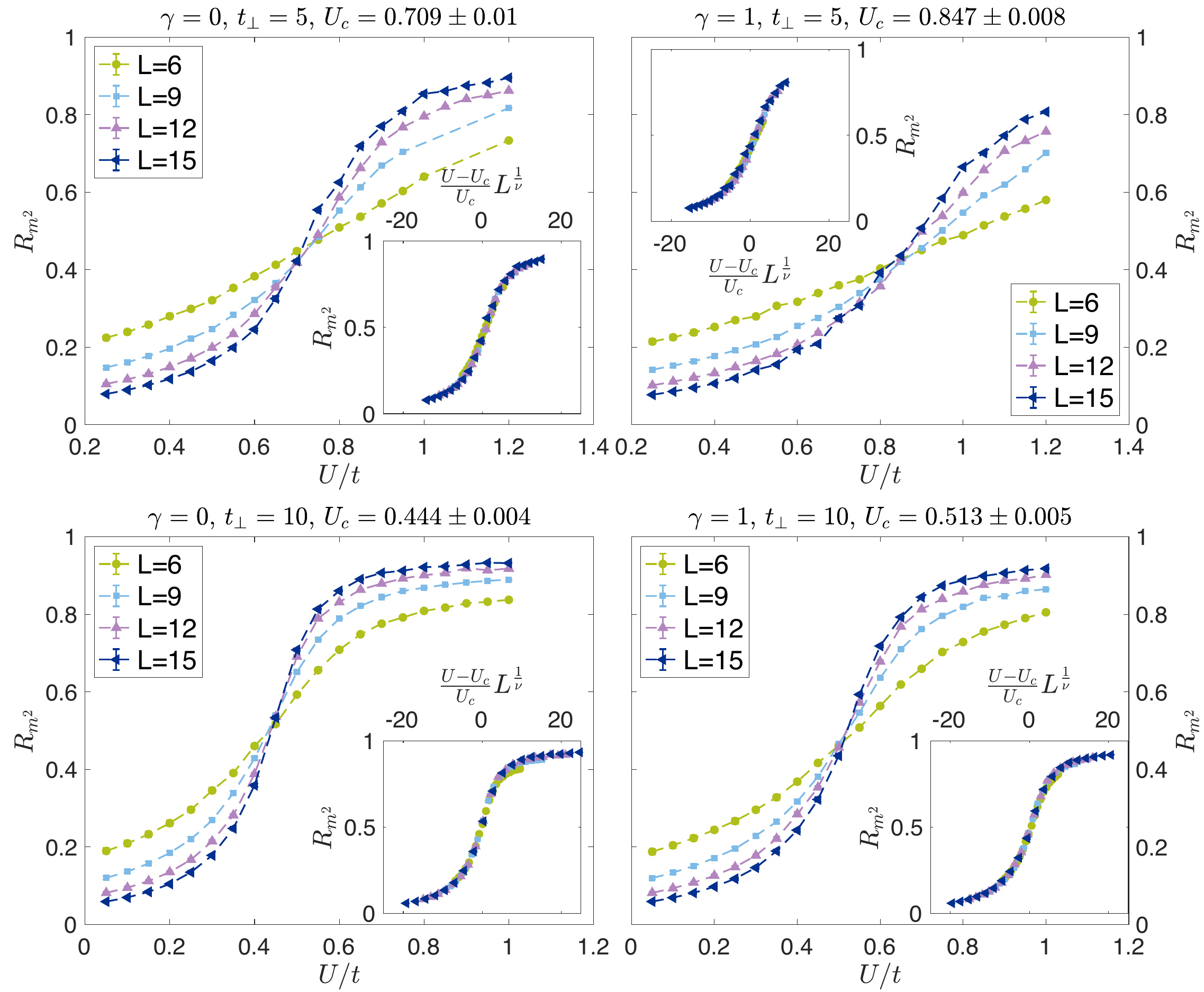}
\caption{Finite size scaling analysis for Hubbard ($\gamma=0$) and Coulomb ($\gamma=1$) interactions with $t_{\perp}=5t,10t$. In the main panels, the correlation ratios $R$ for different system sizes are plotted
against the onsite interaction strength. The crossings of different
data sets show the scale invariance feature, indicating the phase
transition. The interaction strengths at which the crossings occur
are identified as the critical interactions $U_{c}$, with their estimated values listed on top in unit of intralayer hopping $t$. In
the insets, the data sets collapse to a single curve when they are
scaled with the critical exponents $\nu=0.88$, showing that the phase
transition belongs to the appropriate Gross-Neveu universality class~\cite{Herbut2009,Zerf2017,Assaad2012,Toldin14,Otsuka2015}.\label{fig:FiniteSizeScaling}}
\end{figure}

When we approach the strong coupling limit in the Hubbard model, the double occupancy at each site is suppressed,
and the system maps to the $S=1/2$ Heisenberg model. Depending on the ratio of $t_\perp/t$, the system may develop
a long-range Néel order or turn into a dimer phase. The interplay between the antiferromagnetic order and the dimer phase will be presented elsewhere~\cite{Leaw2019}. To study the phase transition, we measure the structure factor that measures both the antiferromagnetic order and the dimer order
$S\left(\mathbf{q}\right)=\frac{1}{2L^{2}}\sum_{ij}e^{i\mathbf{q}\cdot\left(\mathbf{r}_{i}-\mathbf{r}_{j}\right)}\left\langle M_{\text{AF}}\left(\mathbf{r}_{i}\right)\cdot M_{\text{AF}}\left(\mathbf{r}_{j}\right)\right\rangle $,
where $M_{\text{AF}}\left(\mathbf{r}_{i}\right)=\sum_{l}\left[m_{Al}\left(\mathbf{r}_{i}\right)-m_{Bl}\left(\mathbf{r}_{i}\right)\right]$,
and $m_{Cl}=\hat{c}_{i\uparrow l}^{\dagger}\hat{c}_{i\uparrow l}-\hat{c}_{i\downarrow l}^{\dagger}\hat{c}_{i\downarrow l}$
measures the magnetization of the $l$-th layer $C=A,B$ atom in unit
cell positioned at $\mathbf{r}_{i}$. The correlation function $\left\langle M_{\text{AF}}\left(\mathbf{r}_{i}\right)\cdot M_{\text{AF}}\left(\mathbf{r}_{j}\right)\right\rangle $
is computed using the numerically exact zero-temperature projective
quantum Monte Carlo method~\citep{Bercx2017,HokinThesis}, which projects the wave-function to zero temperature
to obtain the ground state properties of the system. The resulting
structure factor $S$ has an anomalous  dimension, which would require an additional exponent when we perform the scaling analysis. We use instead the correlation
ratio $R$ of the structure factor at the ordering momentum $\boldsymbol{\Gamma}$
and the momentum closest to it, 
\begin{equation}
R=1-\frac{S\left(\bm{\Gamma}+\mathbf{b}/L\right)}{S\left(\bm{\Gamma}\right)}
\end{equation}
where $\mathbf{b}$ is the reciprocal lattice vector, and $\mathbf{b}/L$
is the momentum closest to the $\bm{\Gamma}$ point for a system with
size $L\times L$. Since the correlation ratio $R$ is dimensionless,
it scales to 1 in the antiferromagnetic phase and scales to 0 in the
metallic phase (see Fig. \ref{fig:FiniteSizeScaling}). Exactly at
the phase transition, the spin structure factor is scale invariant,
which allows us to determine the critical interaction strength by
pinpointing the Hubbard interaction that shows no change in the correlation
ratio $R$ when the system size is increased.

\begin{figure}
\centering
\subfloat[\label{fig:BilayerPhaseDiagram-a}]{
\includegraphics[width=0.8\linewidth]{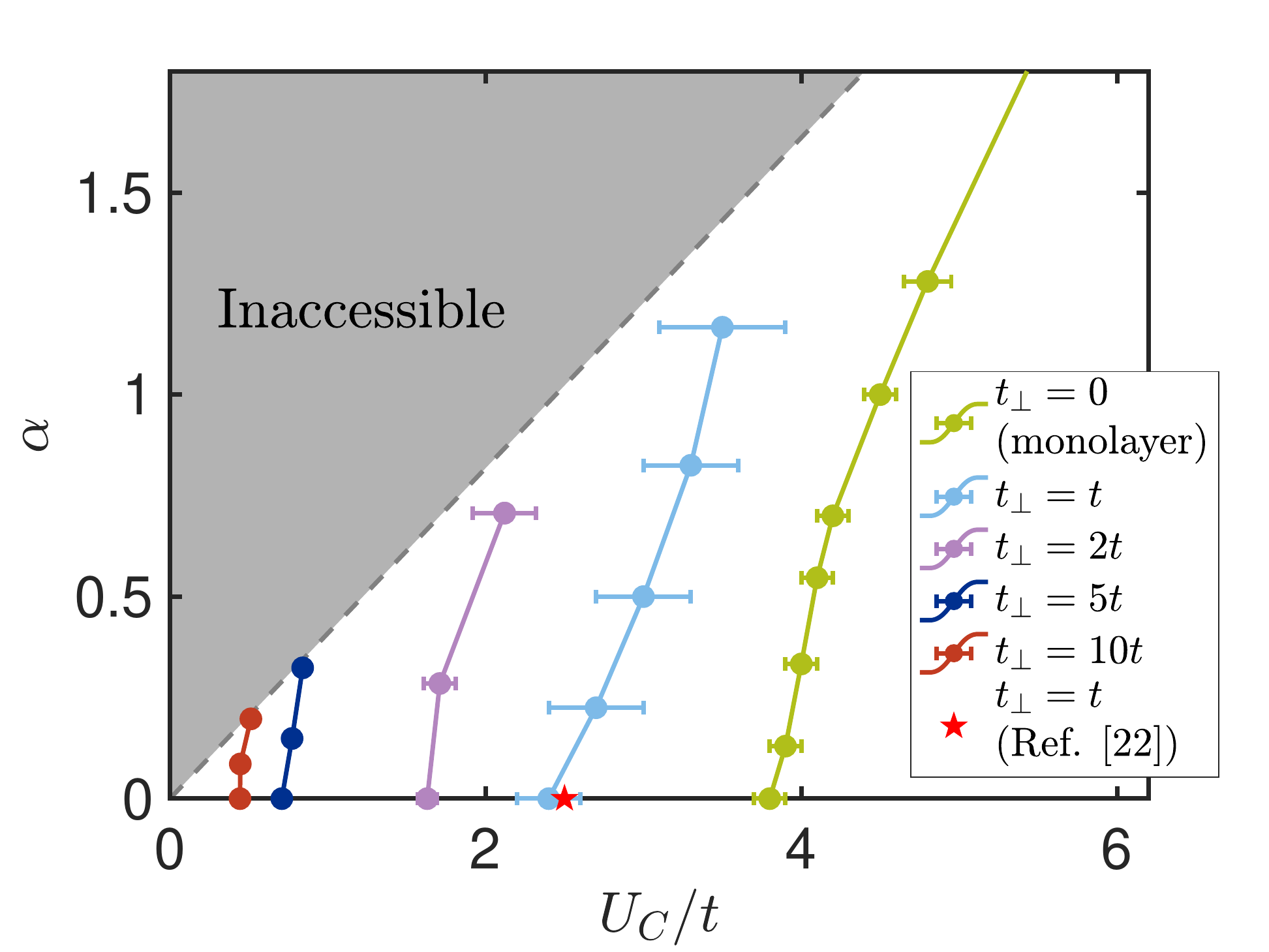}
}\\\subfloat[\label{fig:BilayerPhaseDiagram-b}]{
\includegraphics[width=0.8\linewidth]{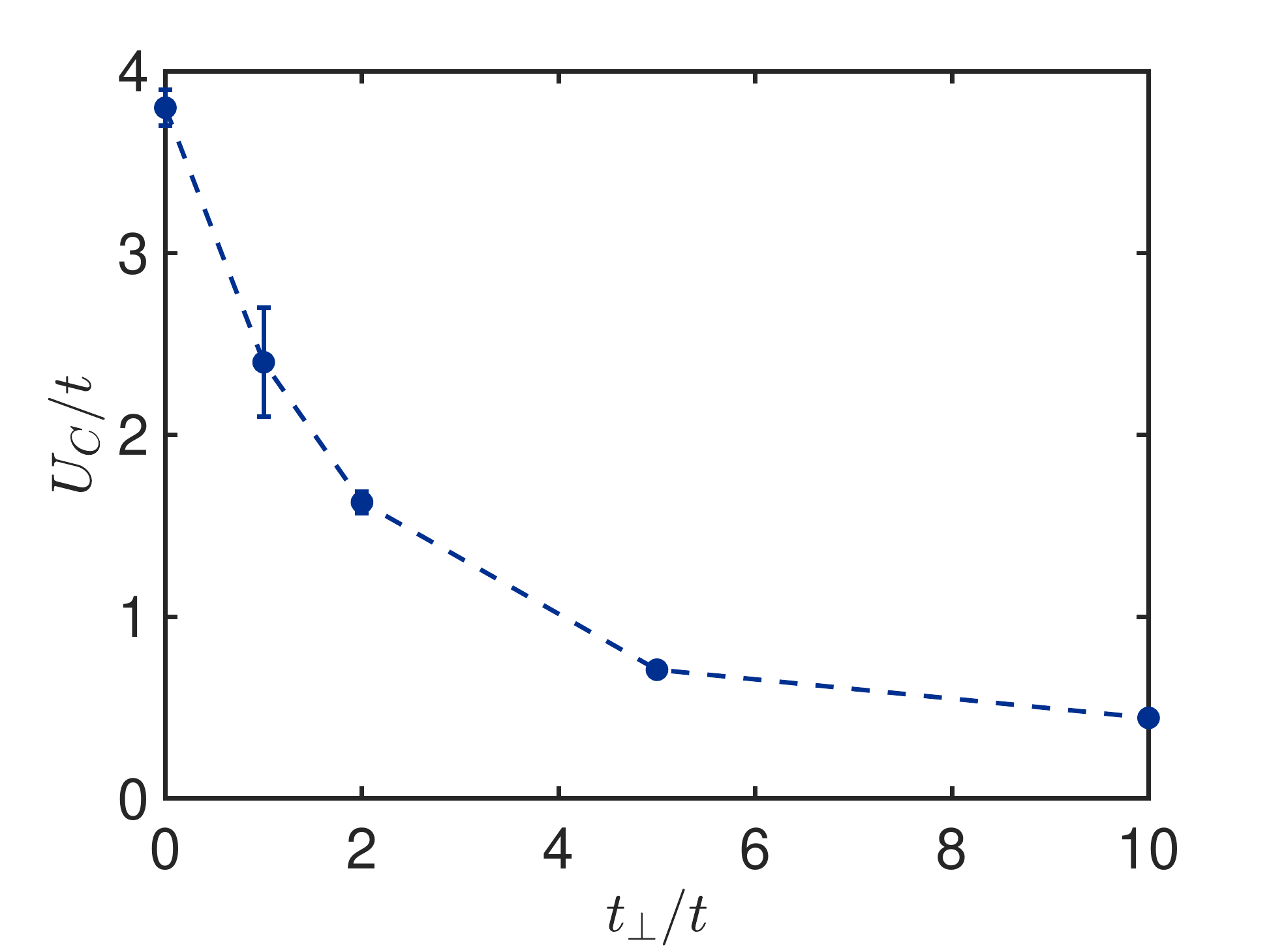}
}
\caption{The critical onsite interaction strengths $U_{c}$ for systems with
different interlayer hopping $t_{\perp}$. (a) The phase boundaries
for different $t_{\perp}$ in the phase space of onsite interaction
$U$ and long-range interaction $\alpha$. The left of the phase boundaries
is the semi-metallic phase, while to the right of the phase boundaries
is the antiferromagnetic Mott insulating phase. The critical onsite
interaction strength $U_{c}$ decreases when we increase the interlayer
hopping $t_{\perp}$ from $t_{\perp}=0$, which is the case of monolayer
graphene, to $t_{\perp}=10t$. The critical onsite interaction strength
$U_{c}$ remains finite and increases when the long-range interaction
$\alpha$ is turned on. The shaded window shows the region inaccessible
to our quantum Monte Carlo method. (b) The critical onsite interaction
strength $U_{c}$ is plotted against the interlayer hopping $t_{\perp}$
for the Hubbard models.
\label{fig:BilayerPhaseDiagram}}
\end{figure}

The zero-temperature projective quantum Monte Carlo method has recently
been adapted to include the long-range Coulomb interaction~\citep{Brower2012,Ulybyshev2013,Hohenadler2014}.
This allows us to study the phase transition in systems with long-range
interactions. We perform the scaling analysis for various values of
$\gamma$, and obtain a critical Hubbard interaction $U_{c}$ for
each value of $\gamma$. Tracing out the $U_{c}$ in the phase space
of onsite and long-range interaction, we obtain the phase boundaries
as shown in Fig. \ref{fig:BilayerPhaseDiagram-a}. We perform this
analysis for various values of $t_{\perp}$, and compare with the
monolayer results \citep{Tang2018}. 

\begin{figure}
\includegraphics[width=1\columnwidth]{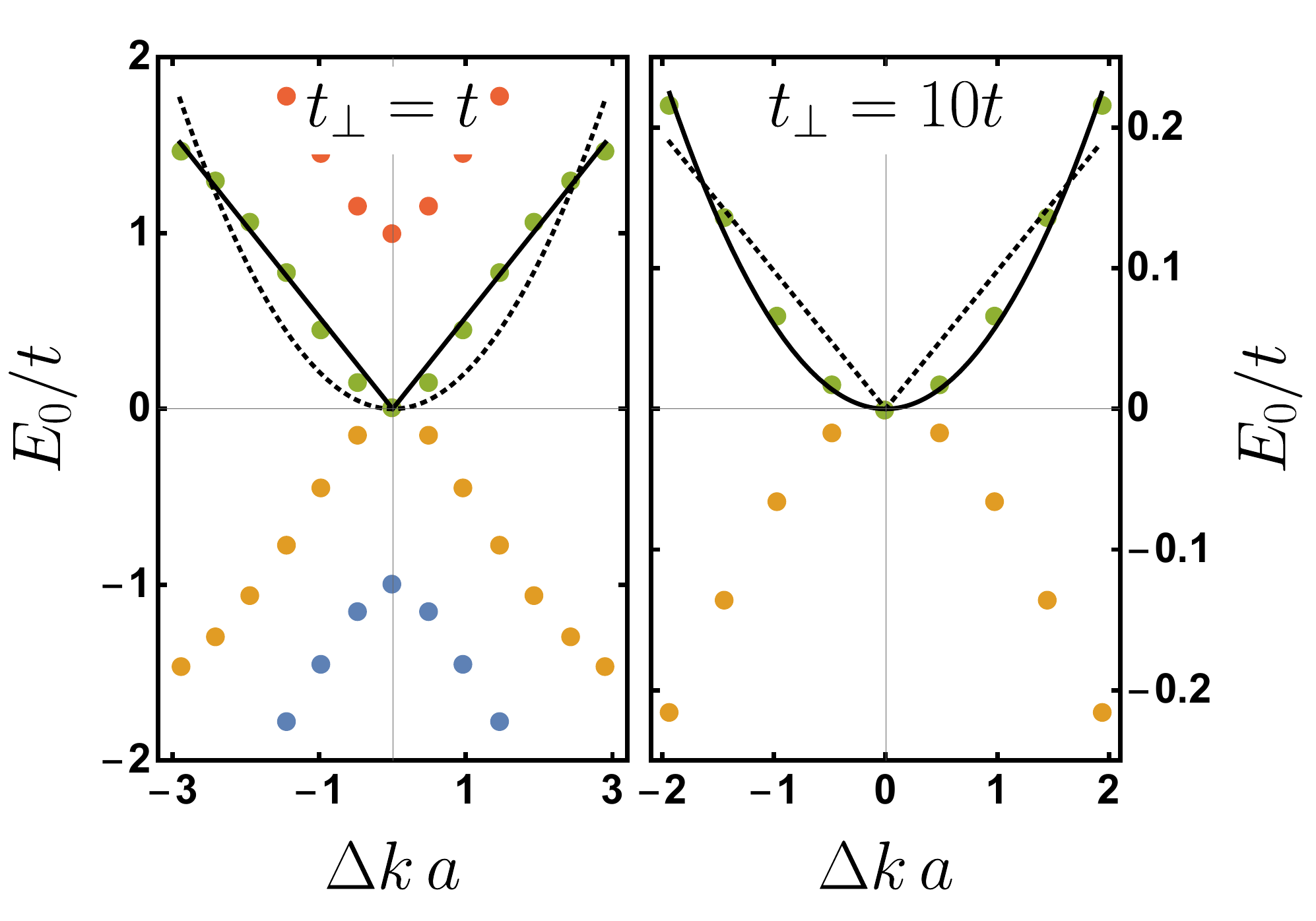}
\caption{The non-interacting spectrum for $t_{\perp}=t$ and $10t$. The data
points show the non-interacting energies at the discrete $k$-points
accessible by the largest system size $L=15$ considered in our quantum
Monte Carlo simulations. The black curves are the linear and quadratic
fits to the points under the curves. For interlayer coupling $t_{\perp}=t$,
the linearity of the non-interacting spectrum is still discernible at these $k$-points, while for large interlayer coupling $t_{\perp}=10t$, the
non-interacting spectrum becomes parabolic. \label{fig:NonInteractingSpectrum}}
\end{figure}

In the canonical Hubbard model, we find that when we increase $t_{\perp}$
from $0$ to $10t$, the critical onsite interaction strength $U_{c}$
decreases gradually from that of the monolayer graphene. The critical
onsite interaction strength $U_{c}$ is plotted against the interlayer
hopping $t_{\perp}/t$ in Fig.~\ref{fig:BilayerPhaseDiagram-b}.
For the case $t_{\perp}=t$, we find $U_{c}\approx3t$ in agreement
with the results in Ref.~\onlinecite{Pujari2016}. However, the critical
value for $t_{\perp}=t$ is close to that of monolayer graphene ($t_{\perp}=0$).
This suggests that for $t_{\perp}=t$, the parabolic part of the low-energy
dispersion is confined to momenta that are inaccessible to the finite
system sizes studied, and instead, the QMC simulations yield results
for the linear part of the spectrum.
%than the critical values of the systems with $t_{\perp}\gg t$. This
%suggests that the interlayer hopping $t_{\perp}=t$ is too weak and
In other words, the system behaves more like a monolayer graphene than a bilayer graphene.
On the other hand, for interlayer hopping as large
as $t_{\perp}=10t$ where the electronic band is parabolic at scales accessible
to the simulations  (see Fig.~\ref{fig:NonInteractingSpectrum}), 
the critical value of the onsite
interaction strength remains small but finite.  In this
limit, the QMC results describe the effects of on-site interactions 
on the parabolic band, and confirm that the a finite, non-zero
critical $U_c$ is required to open up a gap at the Fermi surface. 
This supports the argument
that a linear term is generated to stabilize the Fermi liquid. 

When the long-range interaction is turned on, the critical onsite
interaction strength increases. This can be understood in the following
way. As the linear term is allowed by  symmetry to be dynamically
generated, the long-range interaction also contributes to the linear
term during renormalization group flow. In addition, the mass of the
parabolic term is also reduced by renormalization~\citep{Borghi2009,ViolaKusminskiy2009}.
Both of these amount to the increase in the renormalized energy spectrum,
which decreases the density of states at the Fermi level. Therefore,
a stronger critical onsite interaction is needed to gap out the system.
To show that this is the case, we study the energy spectrum renormalization directly 
in the next section. 

\section{Spectrum renormalization\label{subsec:Spectrum-Renormalization}}

\begin{figure}
\begin{centering}
\includegraphics[width=1\columnwidth]{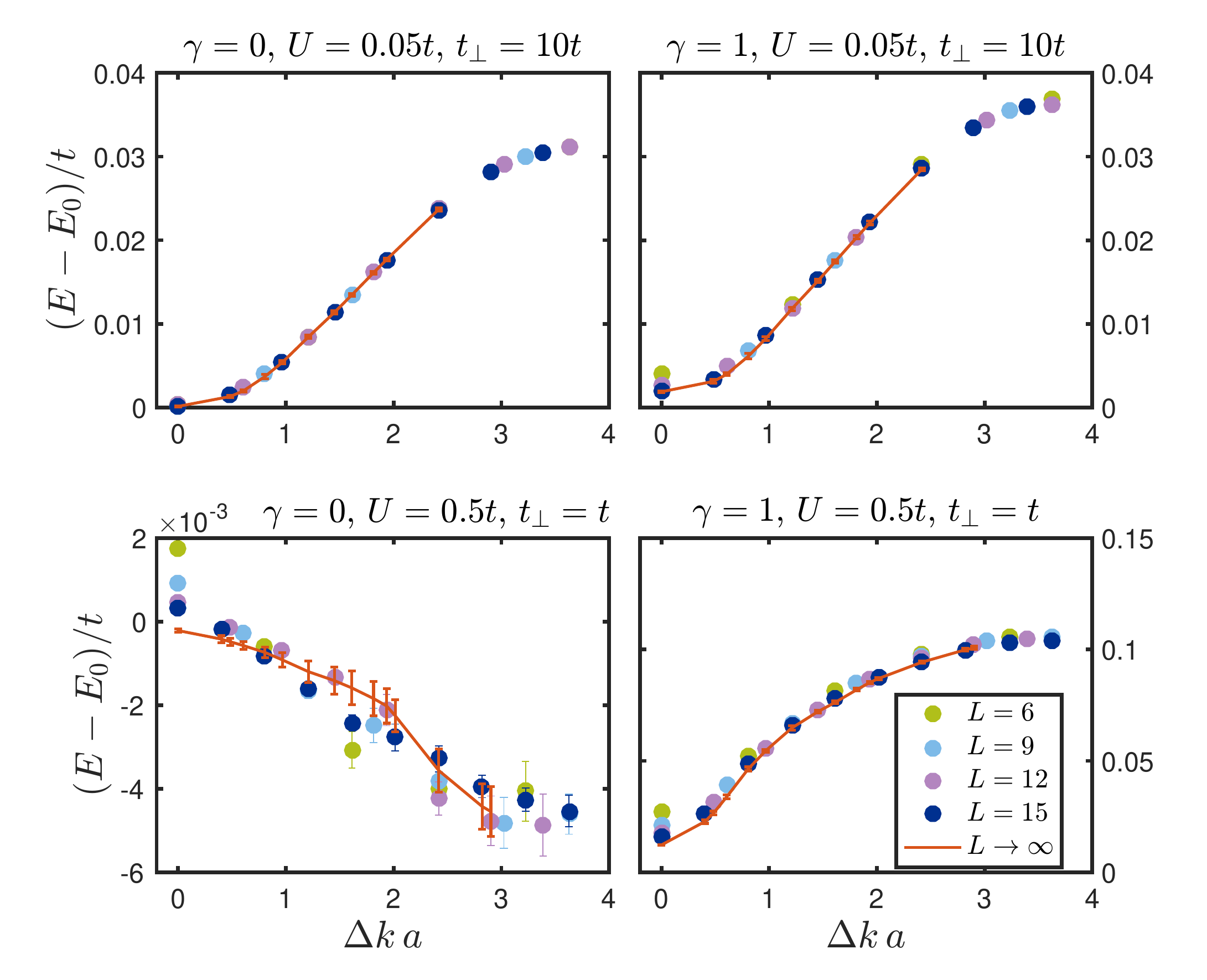}
\par\end{centering}
\caption{The energy renormalization for interacting electrons in bilayer graphene.
The first row shows the energy renormalization $\left(E-E_{0}\right)/t$
of the systems with large interlayer hopping $t_{\perp}=10t$, where
the non-interacting band is expected to be parabolic. In both cases
of purely onsite interaction $\gamma=0$ (top-left panel) and with
the long-range interaction $\gamma=1$ (top-right panel), the energy
renormalizes to a larger value. The second row shows the energy renormalization of
the systems with $t_{\perp}=t$. The energy renormalization shows
similar behavior to that of a monolayer graphene. In the case of Hubbard
model, the energy is renormalized to a smaller value as shown in the
bottom-left panel. With the inclusion of long-range interaction, the
energy is renormalized to a greater value as shown in the bottom-right
panel.\label{fig:EnergyRenormalization}}
\end{figure}

In addition to the phase transition, the zero-temperature projective
quantum Monte Carlo method can also be used to study the spectrum
renormalization, as demonstrated in Ref.~\onlinecite{Assaad2012,Tang2018} for monolayer graphene. Using
the projective quantum Monte Carlo method, we calculate the time-displaced
single-particle imaginary time Green's function $G_{\mathbf{k}}\left(\tau\right)=\sum_{\sigma}\left\langle c_{\mathbf{k},\sigma}^{\dagger}\left(\tau\right)c_{\mathbf{k},\sigma}\left(\tau=0\right)\right\rangle $.
In the limit of $\tau\rightarrow\infty$, the Green's function has
only a single exponential decay $G_{\mathbf{k}}\left(\tau\rightarrow\infty\right)=Z_{\mathbf{k}}e^{-E_{\mathbf{k}}\tau}$,
where $Z_{\mathbf{k}}$ corresponds to the single-particle residue
and $E_{\mathbf{k}}$ is the renormalized single-particle energy.
By fitting our quantum Monte Carlo results at large $\tau$ to an
exponential form, we can extract the renormalized energy $E_{\mathbf{k}}$. 

Fig. \ref{fig:EnergyRenormalization} shows the energy renormalization
$E-E_{0}$ for the Hubbard model $\gamma=0$ and system with the long-range
interaction $\gamma=1$, each for the case of $t_{\perp}=t$ and $t_{\perp}=10t$.
In the system with large interlayer hopping $t_\perp = 10t$, we cannot exclude that a  linear  term is dynamically
generated. Close to the Dirac point the data are consistent with $E(\mathbf{k}) =  v_f| \mathbf{k}|  + | \mathbf{k}|^2/ 2m $   with $ v_f> 0$ such that the density of states at small frequencies reads: $N(\omega) =  \frac{2\pi}{v_f} |\omega| $.  Together with the result that the phase transition for $t_\perp =10t$ occurs at a finite value, quantum Monte Carlo data support the renormalization group argument proposed by Pujari et al~\cite{Pujari2016}.
On the other hand, we see a negative energy renormalization in the
case of the Hubbard model with $t_{\perp}=t$, but a positive energy renormalization
in the case with long-range interaction. A similar trend is shown
in the energy renormalization in monolayer graphene\cite{Tang2018}.
This further supports the view that for  $t_{\perp}=t$ the
QMC simulations on accessible system sizes can only probe the linear part of the 
spectrum and the system behaves
more like a monolayer than a bilayer graphene.

We note that for large interlayer hopping $t_\perp=10t$, the system shows a larger positive energy renormalization in the presence of long-range Coulomb interactions. This indicates that the long-range interactions contribute to generating the linear term, alongside with the contact interactions. Here we show that this is possible within the first-order perturbation theory. Considering the tight-binding model for Bernal-stacked bilayer honeycomb lattice with all the four bands, the non-interacting part of the Lagrangian density is $L_0=\psi^\dagger \mathcal{L}_0 \psi$, with matrix structure\cite{McCann2006} 
\begin{equation}
\mathcal{L}_0=
\begin{pmatrix}-i\omega & 0 & 0 & -t\gamma_{\vec{k}}^{*}\\
0 & -i\omega & -t\gamma_{\vec{k}} & 0\\
0 & -t\gamma_{\vec{k}}^{*} & -i\omega & t_\perp\\
-t\gamma_{\vec{k}} & 0 & t_\perp & -i\omega
\end{pmatrix},
\end{equation}
where $\gamma_{\vec{k}}=\sum_i e^{i\left(\vec{K}+\vec{k}\right)\cdot\vec{\delta}_i}$ is the sum of the phase factors, $\vec{K}=(4\pi/3,0)$ is one of the Dirac points, and $\vec{\delta}_i$ are the positions of the nearest neighbours. In the first-order approximation, the self-energy due to the long-range interaction is 
\begin{equation}
\Sigma(\vec{k})=\int \frac{d\omega\,d^2\vec{q}}{(2\pi)^3}\frac{2\pi e^2}{\left|\vec{k}-\vec{q}\right|}G_0\left(\omega,\vec{q}\right),
\end{equation}
where the non-interacting Green's function is given by the inverse of the Lagrangian density $G_0 = \mathcal{L}_0^{-1}$. After integrating out the frequency $\omega$, we keep the terms that are dominating in the limit of large interlayer hopping $t/t_\perp\rightarrow 0$,
\begin{align}
\Sigma\left(\vec{k}\right)\approx&\frac{e^{2}}{4\pi}\int\frac{q\,dq\,d\theta_{q}}{\sqrt{k^{2}+q^{2}-2kq\cos\left(\theta_{qk}\right)}}\nonumber\\
&\qquad \times \begin{pmatrix}0 & -\frac{\left(\gamma_{\vec{q}}^{*}\right)^{2}}{\left|\gamma_{\vec{q}}\right|^{2}} & 0 & -2\gamma_{\vec{q}}^{*}\frac{t}{t_\perp}\\
-\frac{\gamma_{\vec{q}}^{2}}{\left|\gamma_{\vec{q}}\right|^{2}} & 0 & -2\gamma_{\vec{q}}\frac{t}{t_\perp} & 0\\
0 & -2\gamma_{\vec{q}}^{*}\frac{t}{t_\perp} & 0 & 1\\
-2\gamma_{\vec{q}}\frac{t}{t_\perp} & 0 & 1 & 0
\end{pmatrix},
\end{align}
where $\theta_{qk}\equiv\theta_q-\theta_k$. In the limit of small external momentum $k$, 
\begin{align}
&\frac{q}{\sqrt{k^2+q^2-2kq\cos(\theta_{qk})}}\nonumber\\
&\qquad =1+\frac{\cos(\theta_{qk})}{q}k+\frac{3\cos^2(\theta_{qk})-1}{2q^2}k^2+\mathcal{O}\left(k^3\right)
\end{align}
and approximating $\left(\gamma_{\vec{q}}^*\right)^2/\left|\gamma_{\vec{q}}\right|^2$ to second order in $q$, 
\begin{align}
\frac{\left(\gamma_{\vec{q}}^*\right)^2}{\left|\gamma_{\vec{q}}\right|^2} =& e^{-2i\theta_q}+qe^{-i\theta_q}\frac{e^{6i\theta_q}-1}{4\sqrt{3}}\nonumber\\
&\qquad+q^2e^{2i\theta_q}\frac{e^{6i\theta_q}-1}{48}+\mathcal{O}\left(q^3\right),
\end{align}
we integrate out the angle $\theta_q$ to obtain
\begin{equation}
\Sigma\left(\vec{k}\right)=
\begin{pmatrix}
0 & \alpha & 0 & 0 \\
\alpha^* & 0 & 0 & 0 \\
0 & 0 & 0 & \beta \\
0 & 0 & \beta^* & 0 \\
\end{pmatrix}
\end{equation}
where $\alpha=\alpha_1ke^{-i\theta_k}+\alpha_2\left(ke^{i\theta_k}\right)^2+\mathcal{O}\left(k^3\right)$ and $\beta=\beta_0+\beta_2k^2+\mathcal{O}(k^3)$. Note that only certain terms of $k$ remains, the other terms vanish upon integration over $\theta_q$. The subscripted $\alpha$ and $\beta$ are integrals of $q$,
\begin{align}
\alpha_{1}&=\frac{e^{2}}{4\pi}\int_k^\Lambda dq\,\left[\frac{\pi}{4\sqrt{3}}+\mathcal{O}\left(q^{2}\right)\right]\approx \frac{e^2\Lambda}{4\pi}
\\\alpha_{2}&=\frac{e^{2}}{4\pi}\int_k^\Lambda dq\,\left[-\frac{3\pi}{4q^{2}}+\frac{\pi}{64}+\mathcal{O}\left(q^{2}\right)\right]\approx -\frac{3 e^2}{16 k}
\\\beta_{0}&=\frac{e^{2}}{4\pi}\int_k^\Lambda dq\,2\pi = \frac{e^2(\Lambda-k)}{2}
\\\beta_{2}&=\frac{e^{2}}{4\pi}\int_k^\Lambda dq\,\frac{\pi}{2q^{2}}= \frac{e^2}{8}\left(\frac{1}{k}-\frac{1}{\Lambda}\right).
\end{align}
The integrals are evaluated in the limit of $ka\ll1$ and the ultraviolet cutoff $\Lambda a$ is of the order of 1. In this limit,
$\alpha_{1}$, $\alpha_{2}$,
$\beta_{0}$ and $\beta_{2}$ have some real finite values. We may
now find the renormalized energies by taking the eigenvalues $\epsilon_{n}\left(\vec{k}\right)$
of $\mathcal{H}_{0}\left(\vec{k}\right)+\Sigma\left(\vec{k}\right)$,
where $\mathcal{H}_{0}=\mathcal{L}_{0}+i\omega I_{4}$. Keeping only
terms up to $k^{2}$, we have
\begin{widetext}
\begin{equation}
\mathcal{H}_{0}\left(\vec{k}\right)+\Sigma\left(\vec{k}\right)\approx\begin{pmatrix}0 & \alpha_{1}ke^{-i\theta}+\ensuremath{\alpha}_{2}k^{2}e^{2i\theta} & 0 & \left(\frac{1}{2}\sqrt{3}e^{i\theta}k-\frac{1}{8}e^{-2i\theta}k^{2}\right)t\\
\alpha_{1}ke^{i\theta}+\ensuremath{\alpha}_{2}k^{2}e^{-2i\theta} & 0 & \left(\frac{1}{2}\sqrt{3}e^{-i\theta}k-\frac{1}{8}e^{2i\theta}k^{2}\right)t & 0\\
0 & \left(\frac{1}{2}\sqrt{3}e^{i\theta}k-\frac{1}{8}e^{-2i\theta}k^{2}\right)t & 0 & t'+\beta_{0}+\beta_{2}k^{2}\\
\left(\frac{1}{2}\sqrt{3}e^{-i\theta}k-\frac{1}{8}e^{2i\theta}k^{2}\right)t & 0 & t'+\beta_{0}+\beta_{2}k^{2} & 0
\end{pmatrix}.
\end{equation}
\end{widetext}
The contribution $\sim \beta$ of the self-energy would just renormalize the quadratic bands. However, the contribution $\sim \alpha$ to the self-energy implies that a linear term is generated by the long-range interactions.

\section{Discussion\label{subsec:Discussion}}

Using complimentary RG analysis and large scale unbiased QMC
simulations, we have shown unambiguously that the parabolic low-energy
dispersion of non-interacting electrons on Bernal stacked bilayer honeycomb lattice
is renormalized by the dynamic generation of a linear dispersion due
to the effects of on-site interactions. This linear dispersion stabilizes the
Fermi liquid phase and an interaction-driven
transition to a Mott insulating state occurs at a finite, non-zero critical value
that depends on the magnitude of the inter-layer hopping. Long range
interactions further enhance the linear dispersion and drive the critical
interaction strength to even larger values. Our results further demonstrate that
QMC simulations on bilayer honeycomb lattice needs to be treated with caution. This
is exemplified by the results for the interacting model at small to intermediate
values of the inter-layer hopping ($t_\perp \le t$). While the results yield a 
non-zero critical $U_c$, qualitatively in agreement with expectations from
RG analysis, data for spectrum renormalization at low energies reveal that this
is an artefact of finite system sizes accessible to QMC simulations. The parabolic
dispersion of the non-interacting model is confined to small momenta $ka \lesssim t_\perp/\sqrt{3}t$. For $t_\perp \lesssim t$, to probe the relevant momenta, 
one needs to go to length scales larger than the system sizes that are accessible
to QMC simulations. As a result, the QMC results in this limit correspond to the 
linear (at larger momenta) part of the dispersion and effectively reproduces the
monolayer physics. 

The effects of the dynamically generated linear term in the dispersion can be
probed by QMC simulations at large values of the interlayer coupling $t_\perp \gg t$.
The parabolic dispersion extends to higher momenta that are accessible to available
system sizes in our study and the results are relevant to large length scale physics.

Despite the large discrepancy between the interlayer hopping in the
simulations $t_{\perp}=10t$ and the experimental interlayer hopping
$t_{\perp}\approx0.1t$, we can still glean information from the simulations to understand the experiments.  More specifically, the renormalization group flow filters out the high energy physics to focus on the low energy physics.
During a renormalization
group flow, the high frequency modes are integrated out and the resulting
effective theory is rescaled to be compared with the original theory.
Often in this renormalization group approach, a high energy cutoff
is introduced to specify the validity regime of the theory, where
the effects of renormalization beyond this cutoff are neglected. One
can imagine to start a renormalization group flow from realistic bilayer
graphene parameter $t_{\perp}\approx0.1t$. During the flow only the
interlayer hopping $t_{\perp}$ is rescaled, all other renormalization
effects are neglected. At the point when the interlayer hopping is
rescaled to $t_{\perp}=10t$, we arrive at a new theory for which is cut off at $t_{\perp}=10t$, and that we study using the quantum Monte Carlo simulations.  Although the interaction will also be rescaled to a different value, the conclusion that
the phase transition occurs at finite critical onsite interaction strength remains valid.

Additionally, in realistic bilayer graphene where the interlayer next-nearest-neighbour hopping is allowed, a linear dispersion appears in the momentum scale below the parabolic dispersion. Taken together, both these arguments suggest a finite critical interaction for the transition to a Mott insulating phase in realistic graphene bilayers.

The work was made possible by allocation of computational resources at the CA2DM (Singapore), National Supercomputing Centre in Singapore and the Gauss Centre for Supercomputing (SuperMUC at the  Leibniz Supercomputing Center), and funding by the Singapore Ministry of Education (MOE2017-T2-1-130), Deutsche Forschungsgemeinschaft (SFB 1170 ToCoTronics, project C01) and NSERC of Canada.

\bibliographystyle{apsrev4-1}
\bibliography{library,extralibrary}

\end{document}